%% file: killing-polyoma-5.tex
\documentclass[12pt,notitlepage]{article}
\usepackage{graphicx}
\usepackage{amsmath}
\usepackage{amssymb}
\usepackage{xcolor}
\usepackage[round,sort&compress,comma,numbers]{natbib}

\makeatletter \renewcommand{\@biblabel}[1]{#1.} \makeatother

\usepackage[small,bf]{caption}

\advance\voffset by -1cm \advance\hoffset by -1.2cm
\input commands

\newcommand{\kmax}{\hat k}
\newcommand{\Kmax}{K_{\max}}
\newcommand{\Kmin}{K_{\min}}
\newcommand{\eps}{\epsilon}

\renewcommand{\cite}{\citep}
\renewcommand{\cite}{\citep}

\graphicspath{%
{/home/fly10/vitaly/research/pictures/}%
{figs/}
}

\begin{document}

\title{Similar \vivo killing efficacy of polyoma virus-specific CD8
  T cells during acute and chronic phases of the infection}

\author{{Vitaly V. Ganusov$^1$, Aron E.  Lukacher$^2$, and Anthony M.
    Byers$^3$}
  \\
  {\small $^1$Los Alamos National Laboratory, Theoretical Biology, MS
    K710, Los Alamos, 87545 NM}\\
  {\small $^2$101 Woodruff Circle, School of Medicine,
    Emory University, Atlanta, GA 30322}\\
  {\small $^3$VaxDesign, 12612 Challenger Parkway, Suite 365 Orlando,
    FL 32826 }}

\maketitle

\begin{abstract}
  Viral infections can be broadly divided into infections that are
  cleared from the host (acute) and those that persist (chronic). Why
  some viruses establish chronic infections while other do not is
  poorly understood. One possibility is that the host's immune
  response is impaired during chronic infections and is unable to
  clear the virus from the host.  In this report we use a recently
  proposed framework to estimate the per capita killing efficacy of
  CD8$^+$ T cells, specific for the MT389 epitope of polyoma virus
  (PyV), which establishes a chronic infection in mice. Surprisingly,
  the estimated per cell killing efficacy of MT389-specific effector
  CD8$^+$ T cells during the acute phase of the infection was very
  similar to the previously estimated efficacy of effector CD8$^+$ T
  cells specific to lymphocytic choriomeningitis virus
  (LCMV-Armstrong), which is cleared from the host.  We also find that
  during the chronic phase of the infection the killing efficacy of
  PyV-specific CD8$^+$ T cells was only half of that of cells in the
  acute phase. This decrease in the killing efficacy is again
  surprisingly similar to the change in the killing efficacy of
  LCMV-specific CD8$^+$ T cells from the peak of the response to the
  memory phase. Interestingly, we also find that PyV-specific CD8$^+$
  T cells in the chronic phase of the infection require lower doses of
  antigen to kill a target cell.  In summary, we find little support
  for the hypothesis that persistence of infections is caused by
  inability of the host to mount an efficient immune response, and
  that even in the presence of an efficient CD8$^+$ T cell response,
  some viruses can still establish a persistent infection.

  \vspace{0.2cm}
 
  Abbreviations: PyV, polyoma virus, LCMV, lymphocytic
  choriomeningitis virus, CIs, Confidence intervals, LTR, likelihood
  ratio test

\end{abstract}

\section{Introduction}

CD8$^+$ T cells play an important role in controlling growth of viral
infections both in mice and in humans by multiple mechanisms including
killing of virus-infected cells and producing anti-viral cytokines
\cite{Kaech.nri02,Wherry.jv04}. Following acute viral infections, CD8
T cells follow a defined program of expansion, differentiation and
contraction that results in the clearance of the virus within 1 to 2
weeks of the infection \cite{Kaech.ni01,Badovinac.ni02,DeBoer.ji03}.
During the expansion phase, CD8$^+$ T cells differentiate into
effectors that possess multiple functions and are very efficient
killers \vivo \cite{Regoes.pnas07,Ganusov.jv08,Ganusov.pnas09}.
Although during a chronic viral infection such LCMV clone 13 infection
of mice CD8$^+$ T cells follow a similar program of expansion and
contraction \cite{Althaus.ji07}, control of the virus is limited
\cite{Wherry.jv04}. Persistent viral load progressively leads to
exhaustion of antigen-specific CD8$^+$ T cells which over time lose
various effector functions such as production of IL-2 and TNF-$\alpha$
\cite{Wherry.jv04}. A similar progressive disfunction of
antigen-specific CD8$^+$ T cells has been observed during HIV infection
\cite{Streeck.pm08}.

Why some infections are cleared while others persist is unknown and
likely to be different for different infections. One hypothesis is
that viruses that persist may elicit T cell responses of inadequate
quality that are unable to control viral growth during acute phase of
the infection. Indeed, infection of mice with LCMV clone 13 (that
persists) elicits CD8$^+$ T cell response of a lower magnitude than
during LCMV-Armstrong infection (that is cleared) and this could be a
part of the reason why LCMV clone 13 persists in mice
\cite{Althaus.ji07}.  Another potential reason could be a low effector
to target ratio induced during the infection with the rapidly
replicating LCMV clone 13 \cite{Li.s09}. Finally, a lower efficiency
of virus-specific CD8$^+$ T cells induced during a chronic viral
infection as compared to T cells during acute infections may be
responsible for viral persistence. There have been no studies that
attempted to discriminate between different mechanisms for viral
persistence and to the best of our knowledge there have been no
studies that have measured the \vivo killing efficacy of CD8$^+$ T
cells during chronic viral infections.

Using a recently developed method of \vivo cytotoxicity
\cite{Aichele.i97,Oehen.ji98,Barchet.eji00,Barber.ji03,Byers.ji03}, we
investigate whether infection of mice with polyoma virus (PyV) that
causes a persistent infection elicits a CD8$^+$ T cell response of a
low killing efficacy and whether this killing efficacy changes from
the acute to chronic phase of the infection. Unexpectedly, we found
that the \vivo killing efficacy of PyV-specific CD8$^+$ T cells at the
peak of the immune response is very similar to that of
LCMV-Armstrong-specific CD8$^+$ T cells obtained in our previous study
\cite{Ganusov.pnas09}.  This argues that inability of CD8$^+$ T cell
response to clear PyV infection in the acute phase is not due to a low
killing efficacy of antigen-specific T cells. Furthermore, during the
chronic phase of the infection, PyV-specific CD8$^+$ T cells did not
lose their ability to kill even months after the primary infection,
being only 2 fold less efficient killers than effectors present at the
peak of the immune response.  Thus, even in the presence of an
efficient CD8$^+$ T cell response, PyV is able to establish a
persistent infection in mice, and it is likely that other viral and/or
host factors play in important role in this process.

\section{Material and Methods}

\subsection{Data}

The method of measuring cytotoxicity of CD8$^+$ T cells \vivo has been
described in detail elsewhere (e.g.,
\cite{Ingulli.mmb07,Ganusov.jv08}).  In this report, we reanalyze
recently published data on killing of peptide-pulsed splenocytes by
polyoma virus (PyV) specific CD8$^+$ T cells during acute and chronic
phases of the PyV infection of C3H/HeN mice \cite{Byers.ji03}.  The
reader is referred to the original publication for more detail. In
short, target splenocytes were pulsed with the MT389 peptide of PyV or
were left unpulsed. The peptide MT389 was chosen because it elicits
the dominant CD8$^+$ T cell response in H-2$^k$ (C3H) mice
\cite{Lukacher.ji99}. Cells were transferred into syngeneic mice
either infected with PyV 7 days before (``acute'' phase), or into mice
controlling PyV infection (days 35-175 post infection, ``chronic''
phase).  At day 7 after the infection, the CD8$^+$ T cell response to
PyV reaches its peak, and by 30 to 40 days after the infection, the
CD8$^+$ T cell response largely contracts \cite{Vezys.jem06}. At
different times after the transfer of target cells (15 min, 30 min, 1
hour, 2 hours, and 4 hours), spleens were harvested, and the number of
pulsed and unpulsed targets, splenocytes, and peptide-specific CD8$^+$
T cells was measured.  To investigate how the killing efficacy of
MT389-specific CD8$^+$ T cells may be affected by the peptide
concentration, in a series of experiments target splenocytes were
pulsed with 5 different concentrations of the MT389 peptide (10, 1,
0.1, 0.01, and 0.001 $\mu$M) and killing of peptide-pulsed targets was
measured as described above.

To estimate the per capita killing efficacy of MT389-specific CD8$^+$
T cells we used the average frequency of these cells in spleens of
PyV-infected mice. Our previous analyses suggested that the frequency
of LCMV-specific CD8$^+$ T cells in an individual mouse bears little
information on the number of target cells killed in the same mouse
(Ganusov at al. in preparation). The use of cell frequencies, rather
than cell numbers, is also supported by our previous work
\cite{Ganusov.pnas09} and follows from general physical principles
because killing occurs in a solid organ such as spleen
\cite{Regoes.pnas07}.  In experiments of \citet{Byers.ji03}, using
average frequencies of MT389-specific CD8$^+$ T cells in mice in the
acute or the chronic phase rather than values measured in individual
mice also gave a better fit of the data (results not shown). As in our
previous work \cite{Ganusov.pnas09}, this is likely because
measurements of CTL frequencies were noisy and poorly correlated with
the number of targets killed (results not shown). Therefore, we used
the average frequency of MT389-specific CD8$^+$ T cells in the spleen
for fitting the data. In the acute phase $f=1.13\%$ (with standard
deviation 0.21\%), and in the chronic phase was $f=0.79\%$ (with
standard deviation 0.15\%).

We have previously found that for LCMV-infected mice, the rate of
recruitment of target cells into the spleen was correlated with the
spleen size \cite{Ganusov.jv08}. For PyV-infected mice, a similar
correlation was found although it was not statistically significant
(results not shown). Therefore, to simplify the analysis here we
assumed that recruitment of target cells into the spleen occurs at the
same rate $\sigma$ in all mice (see also below).

\subsection{Mathematical model for the cytotoxicity \vivo assay}

As we have previously described \cite{Ganusov.jv08}, changes in the
number of unpulsed and peptide-pulsed targets in the mouse spleen are
given by the following equations:

\beqa
S(t) &=& {c\over d-\eps}\left[e^{-\eps t}-e^{-dt}\right],\label{eqn:S} \\
T(t) &=& {c\over (\eps+K-d)}\left[e^{-dt}-e^{-(\eps+K)t}
\right],\label{eqn:T}
\eea

\no where $d=\sigma+\eps+\delta$ is the rate of removal of cells from
the blood, $\sigma$ and $\delta$ are the rates of recruitment of
targets from the blood into the spleen or elsewhere, respectively,
$\eps$ is the rate of preparation-induced cell death,
$c=S_B(0)\sigma$, and $S_B(0)=5\times10^6$ is the initial number of
targets adoptively transferred into mice.  Dividing \eref{T} by
(\ref{eqn:S}) we arrive at the experimentally measured ratio of
peptide-pulsed to unpulsed targets, $R(t)=T(t)/S(t)$

\beqa
R(t) &=& {(d-\eps)\over (\eps+K-d)}\left[{e^{-dt}-e^{-(\eps+K)t}\over
    e^{-\eps t}-e^{-dt}} \right].\label{eqn:R}
\eea

This equation can used to estimate the death rate of pulsed targets
$K$ due to CD8$^+$ T cell mediated killing from the changes in the
ratio $R$ over time \cite[see also \fref{killing}]{Ganusov.jv08}. To
estimate the per capita killing efficacy of PyV-specific CD8$^+$ T
cells we let $K=kf$ where $k$ is the per capita killing efficacy and
$f$ is the frequency of MT389-specific CD8$^+$ T cells in the mouse
spleen. Using adoptive transfer of different numbers of LCMV-specific
effectors or memory T cells we have recently shown that killing of
peptide-pulsed targets is proportional to the frequency of effector
CD8$^+$ T cells in the spleen, but killing saturates at high
frequencies of memory CD8$^+$ T cells \cite{Ganusov.pnas09}.  For PyV,
no similar adoptive transfers have been performed, and therefore we do
not know if the death rate of targets saturates with the increasing
frequency of PyV-specific effector and memory CD8$^+$ T cells. This
will be a subject of an additional study. For the following analysis
we thus assume that death rate of peptide-pulsed targets $K$ in the
spleen is proportional to the average frequency of epitope-specific
CD8$^+$ T cells in the acute or chronic phases of the infection. 


In our analysis, we also estimate the maximal ($\Kmax$) and minimal
($\Kmin$) death rate of peptide-pulsed targets due to CD8$^+$ T cell
mediated killing from the \vivo cytotoxicity experiments
\cite{Ganusov.jv08}.  $\Kmax$ is estimated by solving the following
transcendental equation

\beq \Kmax = {(1-{\rm e}^{-\Kmax t})\over R t},\label{eqn:Kmax}\ee

\no and $\Kmin$ is estimated from

\beq \Kmin = -{\log R\over t}.\label{eqn:Kmin} \ee

\no where $R$ is the ratio of the frequency of peptide-pulsed to
unpulsed targets in the spleen at time $t$ after target cell transfer.
The per capita killing efficacy of CD8$^+$ T cells $k$ is then calculated,
assuming the mass-action killing term (see above) by dividing the
death rate of peptide-pulsed targets $K$ by the frequency of
peptide-specific CD8$^+$ T cells in the mouse spleen $f$.

The dependence of the killing efficacy of CD8$^+$ T cells on the peptide
concentration used to pulse target cells was taken in the form of a
Hill function

\beq k = {\kmax p^n \over h^n+p^n}, \label{eqn:hill} \ee

\no where $\kmax$ is the maximal killing efficacy of MT389-specific
CD8$^+$ T cells, $p$ is the peptide concentration used to pulse target
cells, $h$ is the peptide concentration at which killing is
half-maximal, and $n$ is the power of the Hill function.

\subsection{Statistics}

In our previous study \cite{Ganusov.jv08}, we assumed that errors in
measuring the number of unpulsed targets in the spleen and in the
ratio of the frequency of peptide-pulsed to unpulsed targets
peptide-pulsed targets were of a similar magnitude. In PyV-infected
mice this was not the case, since there was a significantly larger
error in estimating the number of targets recruited than the percent
of targets killed (e.g., see \fref{killing}).  Therefore, we used a
general likelihood approach to fit the data on recruitment of targets
to the spleen and killing of peptide-pulsed targets in the spleen (see
Supplementary Information).  Data on recruitment and on killing were
log-transformed but were assumed to have different distribution of
errors (with standard deviations $s_1$ and $s_2$, respectively). In
general this resulted in well-behaved and normally distributed
residuals except of the data when targets were pulsed with a low
peptide concentration.  The latter fits resulted in a skewed
distribution of the residuals (results not shown). Discrimination
between different models was done using either the likelihood ratio
test (LRT) or F-test for nested models \cite{Burnham.b02,Bates.b88}.
Confidence intervals (CIs) were calculated by bootstrapping the data
with 1000 simulations \cite{Efron.b93}.  Fittings were done in
Mathematica 5.2 using the routine \texttt{FindMinimum}.

\section{Results}

\subsection{Estimation of the killing efficacy of polyoma virus
  specific CD8$^+$ T cells }

We have previously developed a mathematical model to estimate the
death rates of peptide-pulsed targets \vivo due to CD8$^+$ T cell
mediated killing and the per capita killing efficacy CD8$^+$ T cells
\cite{Ganusov.jv08,Ganusov.pnas09}. Here, we applied this model to
estimate the per capita efficacy of the immunodominant polyoma virus
(PyV)-specific CD8$^+$ T cell response at killing target splenocytes,
pulsed with MT389 peptide during the acute (day 7) or chronic (days
32-175) phases of the PyV infection of mice \cite{Vezys.jem06}. The
model described the data well (\fref{killing}) and estimated that half
of peptide-pulsed targets are eliminated in 15 minutes in acutely
infected mice and in 47 minutes during the chronic phase of the
infection. Both rates are surprisingly similar to those obtained for
targets expressing peptides from LCMV Armstrong \cite{Ganusov.jv08}.
Even more surprising, the per capita killing efficacy of
MT389-specific effectors ($k\approx 4.16\ min^{-1}$) is very similar
to the estimates obtained in our previous study for effector CD8$^+$ T
cells, specific to the NP396 ($k=5.5\ min^{-1}$) or GP276 ($k=2.35\
min^{-1}$) epitopes of LCMV (see \tref{parameters} and
\cite{Ganusov.pnas09}).  Similarly as in the previous studies
\cite[Ganusov et al. in preparation]{Yates.po07}, we find that the per
capita killing efficacy of CD8$^+$ T cells in the chronic phase of the
infection was only twice as low as the T cell efficacy during the
acute infection (\fref{killing}). Such a change of the killing
efficacy from the peak of the CD8$^+$ T cell response to the chronic phase
was highly significant (LRT: $\chi_1^2=57.11$, $p\ll0.01$).

We found that recruitment of target cells into the spleen occurred at
similar rates in the acute and chronic phases of the PyV infection
(\fref{killing} and \tref{parameters}; $\chi_2^2=1.80$, $p=0.41$).
Approximately 5\% of all unpulsed targets accumulated in the mouse
spleen at 4 hours of the experiments that is consistent with previous
estimates \cite{Blattman.jem02,Ganusov.jv08}. The half-life time of
targets in the blood as determined by the rate of removal of targets
from the blood, $d=\delta+\sigma+\eps$, was estimated at
$T_{1/2}\approx 2$ hours. This estimate is similar to an estimate
obtained in our previous study \cite{Ganusov.jv08}. Interestingly, in
this study we did not find any evidence for the preparation-induced
death rate of targets ($\eps=0$, $\chi_1^2=2.32$, $p=0.13$) that is in
contrast with our previous observation \cite{Ganusov.jv08}. Different
methods of preparing single cell suspension in two laboratories and
type of infection may have contributed to the observed difference in
estimates of this parameter.

\begin{figure}
  \bc
  \includegraphics[width=.95\textwidth]%
  {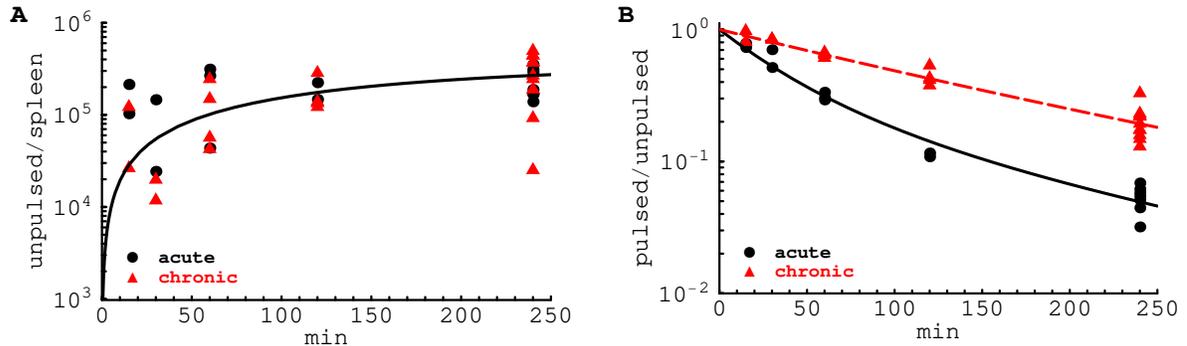}\ec
  \caption{Estimating the efficacy of MT389-specific CD8$^+$ T cells at
    killing targets, pulsed with a high concentration of the peptide
    ($\geq 1\ \mu$M), from the time series data. We fit the model
    prediction on the total number of unpulsed targets in the spleen
    (panel A, see \eref{S}) or the ratio of the frequency of
    peptide-pulsed to unpulsed targets $R$ (panel B, see \eref{R}) to
    the data during acute ($\bullet$) or chronic
    ($\textcolor{red}{\blacktriangle}$) phases of the PyV infection of
    mice.  Parameters providing the best fit of the model to the data
    are given in \tref{parameters}. Recruitment of targets into the
    spleen occurred similarly during the acute and chronic phases of
    the infection ($\chi_2^2=1.80$, $p=0.41$). The half-life time of
    peptide-pulsed targets at the peak of the MT389-specific CD8$^+$ T
    cell response is only 15 minutes, while during the chronic phase,
    half of MT389-expressing targets were eliminated in 47 minutes.
  }\label{fig:killing}
\end{figure}

\begin{table}\bc
  \begin{tabular}{|l|cc|}  \hline
    parameter              & acute          & chronic \\ \hline
    $\sigma,\ 10^{-4}\ min^{-1}$ & \multicolumn{2}{c|}{3.96 (2.90--5.37)}\\ 
    $\delta,\ 10^{-3}\ min^{-1}$ & \multicolumn{2}{c|}{4.92 (2.95--7.04)}\\ 
    $k,\ min^{-1}$            & 4.16 (3.42--4.84) &1.90 (1.65--2.18)\\ \hline
 \end{tabular}
\end{center}
\caption{Parameter estimates obtained by fitting the data on killing
  of targets that were pulsed with a high concentration ($\geq 1\mu
  M$) of the MT389 peptide. Here $\sigma$ and $\delta$ is the rate of
  recruitment of targets from the blood to the spleen or elsewhere,
  respectively, and $k$ is the killing efficacy of MT389-specific CD8
  T cells during acute (left) or chronic (right) phase of the PyV
  infection of mice.  The parameter $\eps$ was fixed at 0 since it did
  not affect the quality of the model fit to data (LRT:
  $\chi_1^2=2.32$, $p=0.13$). The estimated standard deviation of the
  errors in the fit is $s_1 = 0.85\ (0.64-1.04)$ and $s_2=0.19\
  (0.13-0.23)$ (see Materials and Methods for detail). The provided
  95\% CIs were obtained by bootstrapping the data with 1000
  simulations.}
\label{tab:parameters}
\end{table}

To further confirm our estimates of the killing efficacy of
PyV-specific CD8$^+$ T cells, we calculated the minimal and maximal
estimates of the killing efficacy at different times after target cell
transfer using a previously proposed protocol \cite[see also Materials
and Methods]{Ganusov.jv08}.  The estimates of the killing efficacy are
fairly consistent for all time points analyzed for the chronically
infected mice but not for acutely infected mice (\fref{point} in
Supplementary Information). The latter is most likely due to errors
associated with estimating the small number of remaining targets at
the end of the killing assay.  Indeed, linear regression analysis
suggests that the estimate of the killing efficacy is increasing with
the time since cell transfer (results not shown).  Importantly, there
is a considerable overlap between estimates of the killing efficacy of
T cells in the acute and chronic phases of the PyV infection.  This
confirms the analysis of the time series data that on average killing
efficacy of the MT389-specific T cell response in acute phase of the
infection is only two fold lower than that of PyV-specific T cells in
the chronic phase (\fref{killing} and \tref{parameters}). There was no
significant change in the killing efficacy of MT389-specific CD8$^+$ T
cells during the chronic phases of the infection (\fref{k-vs-days})
suggesting that long-term exposure to a persistent infection need not
result in the loss of \vivo effector functions.  This result is
consistent with the previous observation that in chronic PyV
infection, CD8$^+$ T cells are mainly maintained by the production of
effector T cells from newly recruited naive PyV-specific T cells
\cite{Vezys.jem06}.  Nevertheless, this contrasts with the data on
chronic LCMV infection during which LCMV-specific CD8$^+$ T cells lose
their effector functions over time \cite{Wherry.jv03}.

\begin{figure}
  \bc
  \includegraphics[width=0.5\textwidth]{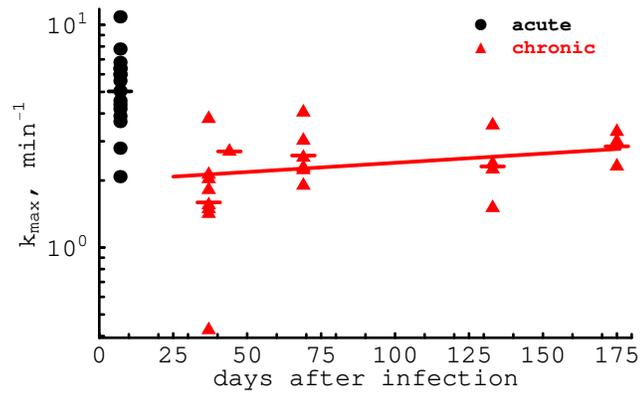}
  \ec
  \caption{The per capita killing efficacy of PyV-specific CD8$^+$ T
    cells does not change during the chronic phase of the infection.
    We estimate the killing efficacy of MT389-specific CD8$^+$ T cells
    from the cytotoxicity assay by normalizing the death rate of
    peptide-pulsed targets $\Kmax$ by the average frequency of
    epitope-specific CD8$^+$ T cells in the spleen (see Materials and
    Methods). Bars show the average estimate per time point.
    Regression slope is not significantly different from zero (F-test,
    $p=0.17$ with one outlier with a low estimate for $k_{\max}$
    removed). The independence of the killing efficacy of CD8$^+$ T cells
    in the chronic phase of the infection from the time since
    infection was further confirmed by fitting longitudinal killing
    data (as in \fref{killing}B) as the fit was not significantly
    improved by assuming that killing efficacy was different for
    different days during the chronic phase of the infection (F-test
    for nested models, $F_{4,16}=2.06$,
    $p=0.13$).}\label{fig:k-vs-days}
\end{figure}

\subsection{Killing efficacy at lower peptide concentrations}

It has been known that the number of targets killed in the \vivo
cytotoxicity assay depends strongly on the amount of peptide used to
pulse target cells (e.g., \cite{Byers.ji03}). However, it has not been
investigated how this translates into the killing efficacy of CD8$^+$
T cells and whether there is a difference in the killing efficacy of T
cells over the course of a chronic infection at low peptide
concentrations.

To investigate how the killing efficacy of CD8$^+$ T cells depends on
the concentration of the peptide used to pulse target cells, we use
two approaches.  We estimated the per capita killing efficacy of
MT389-specific T cells for every mouse from single measurement of
killing using \eref{Kmax} and (\ref{eqn:Kmin}) (see Materials and
Methods). As expected, we found that the killing efficacy of CD8$^+$ T
cells was higher when targets were pulsed with a higher peptide
concentration (\fref{peptide-log}).  Surprisingly, we find little
difference in the killing efficacy of MT389-specific CD8$^+$ T cells
in the acute and chronic phases of the infection when the peptide
concentration was low. In fact, at the lowest peptide concentration
($0.001\ \mu$M), MT389-specific CD8$^+$ T cells in the chronic phase
appeared to be more efficient at clearing peptide-pulsed targets than
PyV-specific CD8$^+$ T cells in the acute phase of the infection
(\fref{peptide-log}).  In contrast, at high peptide concentrations, T
cells in the acute phase were more efficient at killing targets than
PyV-specific T cells in the chronic phase as expected from the above
analysis (e.g., see \tref{parameters}).

To verify these predictions we refitted the time series data on
killing of targets pulsed with peptides at different concentrations
assuming that the killing efficacy of CD8$^+$ T cells depends on the
amount of peptide that target cells present (see \eref{hill} in
Materials and Methods). The model described the data reasonably well
(\fref{killing-peptide}) with the exception of the data with peptide
concentration $0.01\ \mu$M where the model underestimated the amount
of killing (see also \fref{peptide-log}).  Interestingly, the model
fits also predicted a lower half-saturation constant $h$ (i.e., a
higher affinity to the antigen) of MT389-specific CD8$^+$ T cells in
the chronic phase of the infection than that of cells in the acute
phase (\tref{parameters.peptide} and \fref{k_vs_peptide} in
Supplementary Information). The fits also confirmed a higher killing
efficacy of CD8$^+$ T cells in the acute phase at high concentrations of
the MT389 peptide (\tref{parameters.peptide}).

\begin{figure}
\bc
\includegraphics[width=.5\textwidth]{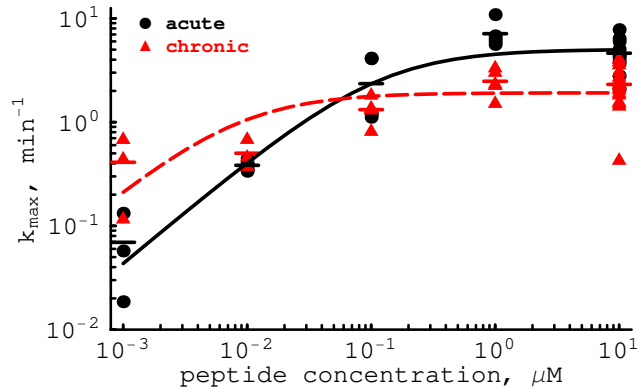}
\ec
\caption{Estimating the killing efficacy of the PyV-specific CD8$^+$ T
  cells in acute and chronic phases of the infection from single
  measurements of killing.  Points represent the maximal killing
  efficacy of MT389-specific CD8$^+$ T cells, estimated for individual
  mice, and horizontal lines are averages for a given peptide
  concentration.  Solid lines show the fits of a Hill function given
  in \eref{hill} with $n=1$ to the estimated killing efficacies.  The
  estimated parameters are: $\kmax=4.34\ min^{-1}$, $h = 0.12\ \mu M$
  (acute) and $\kmax=1.91\ min^{-1}$, $h = 0.008\ \mu M$ (chronic).
  Differences in $\kmax$ and $h$ for cells in the acute and chronic
  phases of the infection remained if we used the minimal estimate of
  the killing efficacy (results not shown)}\label{fig:peptide-log}
\end{figure}

\begin{figure}
  \bc
  \includegraphics[width=\textwidth]{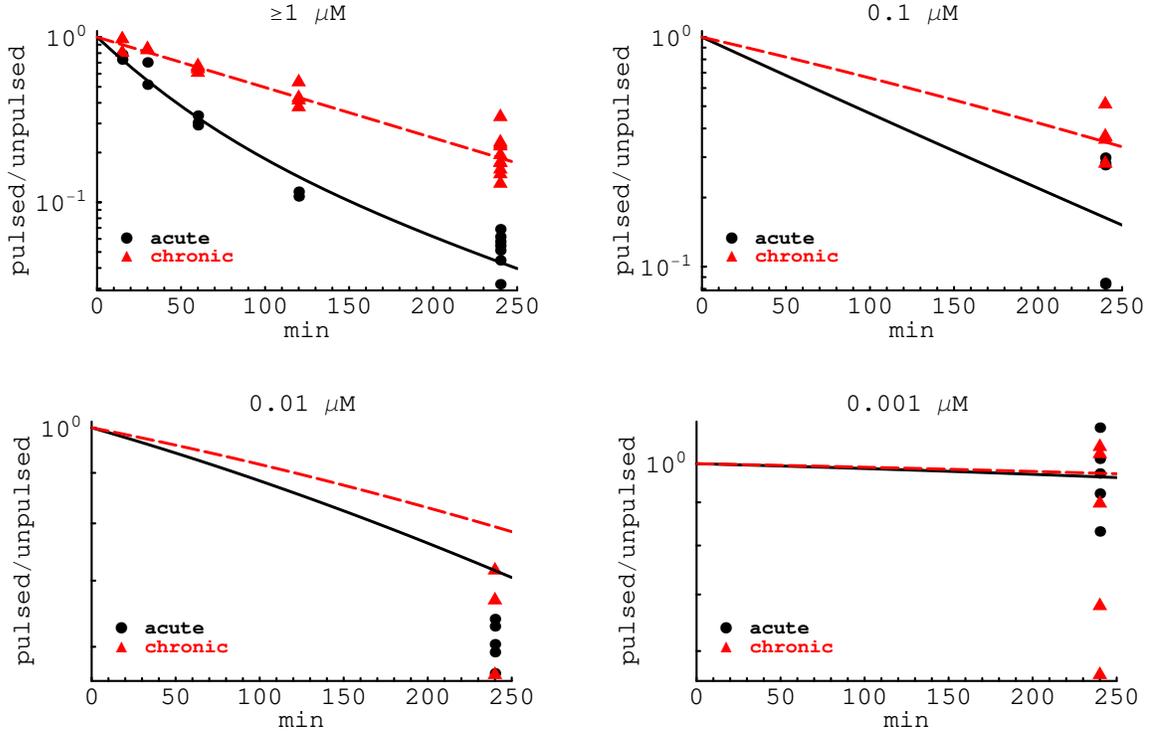}
\ec
\caption{Estimating the efficacy of PyV-specific CD8$^+$ T cells at
  killing target cells, pulsed with different concentrations of
  peptide MT389, from the time series data. Points are the measured
  ratios of the frequency of peptide-pulsed to unpulsed target cells,
  and lines show the fits of the model prediction to the data.  The
  data were fitted simultaneously assuming that the dependence of the
  killing efficacy on the peptide concentration follows a Hill
  function (see \eref{hill}). Parameters providing the best fit of the
  model to data are given in \tref{parameters.peptide}. Note that to
  show the fits and the data, we used different scales on $y$-axes.
}\label{fig:killing-peptide}
\end{figure}

\begin{table}
\bc
  \begin{tabular}{|l|cc|}  \hline
    parameter              & acute          & chronic \\ \hline
    $d,\ 10^{-3}\ min^{-1}$ & \multicolumn{2}{c|}{5.83 (3.61--8.78)}\\ 
    $\kmax,\ min^{-1}$        & 3.50 (2.93--4.04) &1.89 (1.58--2.16)\\
    $h,\ \mu M$            &0.19 (0.10--0.37)&0.08 (0.04--0.14)\\ \hline
\end{tabular}
\end{center}
\caption{Parameter estimates obtained by fitting the data on killing
  of targets that were pulsed with different concentrations of the
  MT389 peptide.  The data were fitted as described in the legend of
  \fref{killing-peptide}, assuming different maximal killing
  efficacies during acute and chronic phases of the infection
  ($\kmax$) and different half-saturation constants ($h$, see
  \eref{hill} in Materials and Methods).  Importantly, the
  half-saturation constant was different for CD8$^+$ T cells in acute and
  chronic PyV infection (F-test for nested models, $F_{1,61}=4.7$,
  $p=0.034$). The shape parameter of the Hill function $n$ was not
  significantly different from 1 (F-test for nested models,
  $F_{1,60}=3.57$, $p=0.064$). }\label{tab:parameters.peptide}
\end{table}

\section{Discussion}

There is a limited understanding of why CD8$^+$ T cell responses in
some cases can control and clear viral infections while in other cases
they fail to do so. It was recently shown that the initial effector to
target ratio during acute phase of LCMV infection may determine
whether the infection is cleared (high E/T ratio) or persists (low E/T
ratio, \cite{Li.s09}).  Our results demonstrate that in the case of
PyV infection of mice, the failure to clear the infection is not due
to a reduced killing efficacy of PyV-specific CD8$^+$ T cells.
Instead we found that killing efficacy of PyV-specific effectors is
very similar to that of LCMV-specific CD8$^+$ T cells that clear the
infection from mice within 7 days \cite{Lau.n94}. Other mechanisms
such as an ability to persist in a silent form by expressing proteins
at a low level, may have allowed PyV to escape the eradication from
infected mice. It will be interesting to investigate if infection of
mice with other strains of LCMV, some of which induce high levels of
IL-10 \cite{Ejrnaes.jem06,Brooks.nm06,Maris.bi07}, establish
persistence by reducing the killing efficacy of LCMV-specific CD8$^+$
T cells and not necessarily due to a low ratio of effectors to targets
\cite{Li.s09}.

We found that PyV-specific CD8$^+$ T cells during the chronic phase of
the infection have a lower half-saturation constant than T cells
present at the peak of the immune response (\tref{parameters.peptide}
and \fref{k_vs_peptide} in Supplementary Information). A lower
half-saturation constant suggests that these CD8$^+$ T cells are more
sensitive to lower quantities of the antigen than effectors.  The
effect of downregulation of T cell receptors on T cells, thereby
reducing TCR avidity to the antigen, (``tuning of activation
threshold'') has been proposed to occur in some situations, for
example, during maturation of T cells in the thymus
\cite{Grossman.pnas92,Grossman.pnas96}, and has been documented to
occur during acute infections \cite{Slifka.ni01,Zehn.n09}.  Our
results indicate that CD8$^+$ T cells may also regulate their avidity
for the antigen when antigen load is low. In the acute phase, when
antigen is abundant, CD8$^+$ T cells with lower avidity are selected.
In the chronic phase, when antigen is limited, to efficiently control
virus replication CD8$^+$ T cells with higher avidity may become
dominant.  Such a selection could occur via at least two mechanisms.
First, higher avidity T cells may be selected during the chronic phase
of the infection from the existing pool of effector T cells.
Alternatively, PyV-specific naive CD8$^+$ T cells that have recently
emigrated from the thymus and have a high affinity for MT389 epitope
of PyV could be preferentially activated by the low amount of antigen
present during the infection \cite{Vezys.jem06}. Given the inability
of PyV-specific CD8$^+$ T cells to self-renew during the chronic phase of
the infection, the latter mechanism seems more likely. One potential
way of regulating avidity for the antigen is the amount of T-cell
receptors present on the cell surface. A testable prediction of this
model result would be a higher number of T-cell receptors on
PyV-specific CD8$^+$ T cells during the chronic phase of the infection
as compared to effectors at the peak of the immune response.

In contrast to the substantial body of work characterizing memory CD8
T cell exhaustion in high-load chronic viral infections, we know
surprisingly little about maintenance of memory CD8$^+$ T cells to
viruses that establish low-level persistent infection. Persistent
viremia (i.e., high-load persistent infection) is associated with a
number of infections of global import, including HIV, HCV, and HBV. In
mice, infection by high replicating strains of LCMV (e.g., clone 13 or
DOCILE) has been the principal animal model for high-load persistent
infection. In this setting, LCMV-specific CD8$^+$ T cells
progressively accumulate defects in effector function and replication
potential and are then deleted, a phenomenon called ``exhaustion''
\cite{Zajac.jem98,Wherry.jv04,Wherry.i07}. The high antigenic burden
and duration associated, for example, with HIV infection also drive
virus-specific CD8$^+$ T cells to exhaustion, which is generally
considered to be a prelude to viral persistence
\cite{Streeck.pm08,Rowland-Jones.pm08}.  T cell exhaustion, however,
is at odds to situations involving low-load persistent viral
infections (e.g., $\gamma$HV-68, and PyV in mice) where the majority
of CD8$^+$ T cell functions as measured in ex vivo assays are not
compromised and stable numbers of virus-specific T cells are
maintained \cite{Kemball.ji05,Obar.jv06,Cush.ji07,Cush.ji09}.  Here we
further demonstrate that the \vivo killing efficacy of PyV-specific
CD8$^+$ T cells in the chronic phase of the infection is only two-fold
lower than that of cells in the acute phase arguing that CD8$^+$ T
cells can maintain their \vivo effector functions long since the
primary phase of a persistent infection (\fref{k-vs-days}).

In our previous study by adoptively transferring different numbers of
LCMV-specific effector and memory CD8$^+$ T cells, we found that per cell
killing efficacy of memory cells decreased with the number of
transferred cells. We conjectured that this effect could arise due to
competition between LCMV-specific memory CD8$^+$ T cells in the white pulp
of the spleen for the access to targets \cite{Ganusov.pnas09}. For PyV
infection, we do not know if the two fold reduced killing efficacy of
CD8$^+$ T cells in the chronic phase of the infection is also due to
localization of these cells in the white pulp of the spleen.

We have found that killing efficacy of PyV-specific CD8$^+$ T cells
depended strongly on the peptide concentration used to pulse targets.
To kill a peptide-pulsed target, a CD8$^+$ T cell has to accomplish
two tasks: find the target and deliver the lethal hit. It is unclear
which of these two processes is the limiting step in the killing of
targets.  Additional studies including those looking at the process of
killing of targets by effector T cells \vivo (e.g., \cite{Mempel.i06})
will be necessary to investigate this issue further.

Overall, our analysis suggests that PyV-specific CD8$^+$ T cells are
highly efficient \vivo killers of peptide-pulsed targets. However, it
is unknown if these T cells are also good killers of cells naturally
infected with PyV. Also, it is not known if the killing efficacy of
PyV-specific T cells depends on the tissue in which killing is
measured.  Finally, by combining the measured dynamics of viral load
in tissues, PyV-specific T cell response and killing efficacy of T
cells, it will be important to investigate whether CD8$^+$ T cells are
mainly responsible for the decline on PyV load during acute phase of
the infection.  Addressing these issues will a subject of future
research.

\section{Acknowledgments}

This work was supported by the Marie Curie Incoming International
Fellowship (FP6, VVG), U.S.  Department of Energy through the
LANL/LDRD Program (VVG) and NIH (R01CA71971, AEL).

\bibliography{/home/fly10/vitaly/refs/bibliography/library-main}

\newpage
\listoffigures

\newpage
\section{Supplementary Information}

\setcounter{equation}{0}
\renewcommand{\theequation}{A.\arabic{equation}}
\setcounter{figure}{0}
\renewcommand{\thefigure}{S\arabic{figure}}

\subsection{General likelihood approach for data fitting}

When a given model is fitted to several datasets simultaneously, it is
often assumed that the distribution of errors of the fit are similar
for all datasets. This needs not be true in general. In those cases,
when this is not expected to be a correct assumption, a general
likelihood approach should be used. For our model and the data, the
general likelihood of observing the data given the model prediction on
the number of recruited targets $S(t)$ and the ratio of
peptide-pulsed to unpulsed targets $R(t)$ is

\beq L(\vec{S},\vec{R}|\vec{p}) = \prod_{i=1}^n {1\over \sqrt{2\pi
  s_1^2}}{\rm e}^{{-(\log S_i-\log S(t_i))^2\over
    2s_1^2}}\times\prod_{i=1}^n {1\over \sqrt{2\pi s_2^2}} {\rm e}^{{-(\log
    R_i-\log R(t_i))^2\over 2s_2^2}},\label{eqn:likelihood-1}\ee

\no where $S_i$ and $S(t_i)$ are the measured and predicted number of
unpulsed targets in the spleen at time $t_i$ after cell transfer,
respectively, and $R_i$ and $R(t_i)$ are the measured and predicted
ratio of peptide-pulsed to unpulsed targets in the spleen at time
$t_i$ after cell transfer, respectively, $s_1$ and $s_2$ are the
standard deviation of the errors in the recruitment and killing data,
respectively, $n$ is the number of measurements, and $\vec{p}$ is the
vector of model parameters to be estimated from the data. Note that in
this example we have log-transformed the data and the model prediction
for both datasets, and in general, one could use different
transformations for data fitting. 

It is convenient to rewrite \eref{likelihood-1} in terms of
log-likelihood $\mathcal{L} = \log(L)$:

\beq \mathcal{L}(\vec{S},\vec{R}|\vec{p}) = -\sum_{i=1}^n {(\log
  S_i-\log S(t_i))^2\over 2s_1^2}-\sum_{i=1}^n {(\log R_i-\log
  R(t_i))^2\over 2s_2^2}-n\log(s_1 s_2),\label{eqn:likelihood-2}\ee

\no where we omitted the constant term $-n\log(2\pi)$. Parameters of
the model are found by maximizing the log-likelihood in respect to
model parameters and error variances $s_1^2$ and $s_2^2$. In the case
when $s_1=s_2$, maximization of \eref{likelihood-2} will be identical
to the commonly used method of least squares \cite{Burnham.b02}.

\begin{figure}[p]
  \bc
  \includegraphics[width=.5\textwidth]{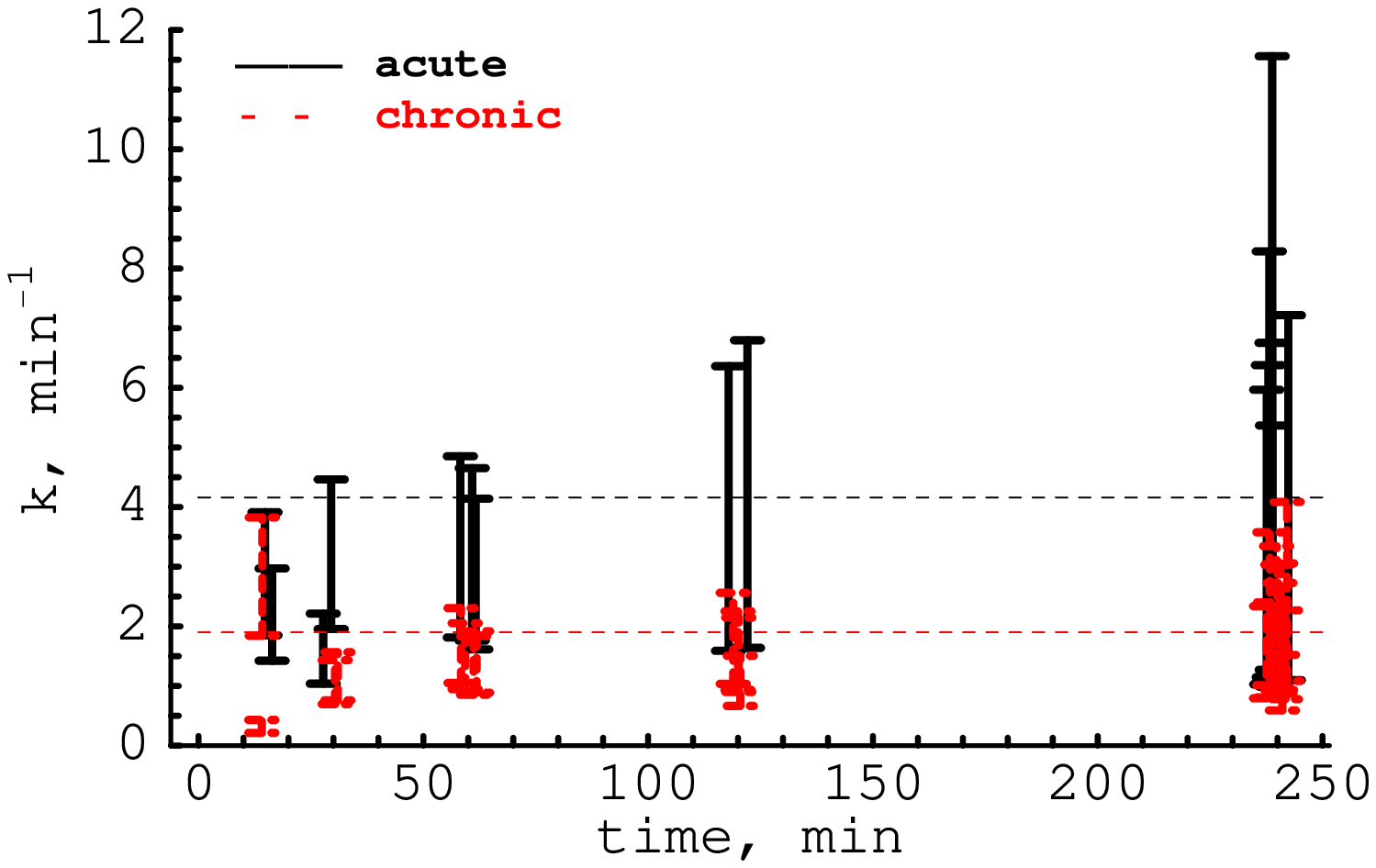}\ec
  \caption{Estimating killing efficacy of MT389-specific CD8$^+$ T
    cell from a single measurement of \vivo cytotoxicity.  Using
    simple formulas (see Materials and Methods) we estimate the range
    (minimal and maximal estimates) for the efficacy of CD8$^+$ T
    cells at killing targets pulsed with a high concentration of the
    MT389 peptide polyoma virus ($\geq 1\ \mu$M).  Mean and standard
    deviations are $k_{\max} = 5.74\pm 2.24\ min^{-1}$ and $k_{\min} =
    1.40\pm 0.35\ min^{-1}$ (acute phase), $k_{\max} = 2.37\pm 0.87 \
    min^{-1}$ and $k_{\min} = 0.89\pm 0.29\ min^{-1}$ (chronic phase).
    Horizontal dashed lines denote the killing efficacy of CD8$^+$ T
    cells as estimated from the time series data (see
    \tref{parameters} in the Main text). }\label{fig:point}
\end{figure}

\begin{figure}
  \bc
  \includegraphics[width=0.5\textwidth]%
{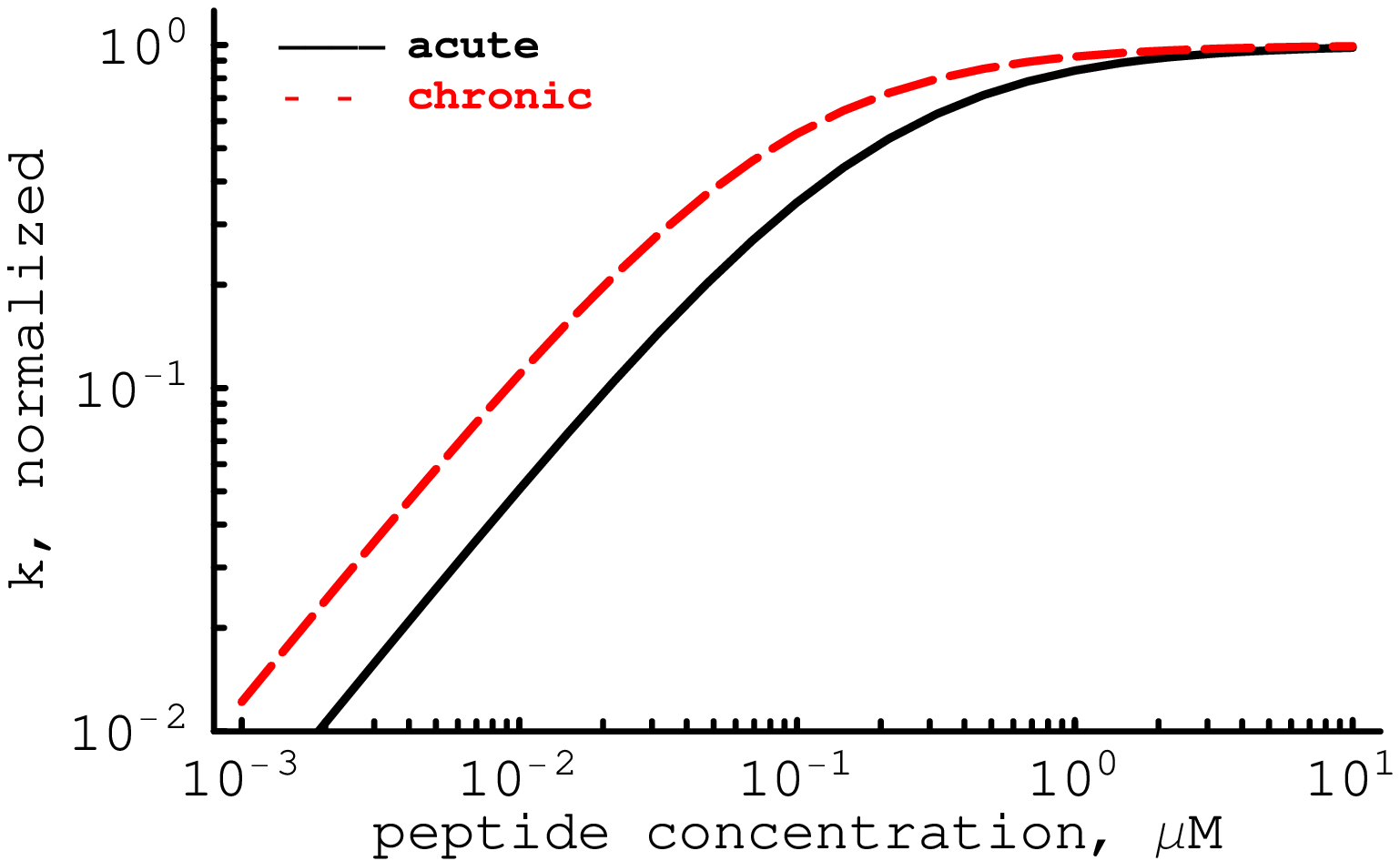}
  \ec
  \caption{Predicted changes in the normalized killing efficacy of
    MT389-specific CD8$^+$ T cells in acute (solid lines) or chronic
    (dashed line) phases of PyV infection. Killing efficacy is
    normalized to its maximal value. PyV-specific CD8$^+$ T cells in
    the chronic phase of the infection are more efficient killers at
    low peptide concentrations. Parameters estimates are given in
    \tref{parameters.peptide}.}\label{fig:k_vs_peptide}
\end{figure}

\end{document}

%% file: commands.tex
\newcommand{\bii}{\begin{itemize}}
\newcommand{\eii}{\end{itemize}}
\newcommand{\bie}{\begin{enumerate}}
\newcommand{\eie}{\end{enumerate}}
\newcommand{\beq}{\begin{equation}}
\newcommand{\ee}{\end{equation}}
\newcommand{\eeq}{\end{equation}}
\newcommand{\beqa}{\begin{eqnarray}}
\newcommand{\eea}{\end{eqnarray}}

\newcommand{\no}{\noindent}

\newcommand{\bc}{\begin{center}}
\newcommand{\ec}{\end{center}}

\newcommand{\fref}[1]{Figure\ \ref{fig:#1}}
\newcommand{\tref}[1]{Table\ \ref{tab:#1}}

\newcommand{\eref}[1]{eqn.\ (\ref{eqn:#1})}

\newcommand{\vivo}{{\it in vivo\ }}

\newcommand{\mat}{\begin{array}{cc}}
\newcommand{\emat}{\end{array}}

\newcommand{\LD}{\LD_{50}}